\documentclass[twocolumn,twoside,slac]{revtex4}
\usepackage{graphicx}
\usepackage{fancyhdr}
\begin{document}

\title{McRunjob: A High Energy Physics Workflow Planner for Grid Production
                              Processing}

%

\author{G.E. Graham}
\affiliation{Fermi National Accelerator Laboratory, Batavia, IL, 60510-0500, USA}
\author{D. Evans and I. Bertram}
\affiliation{Lancaster University, Bailrigg, Lancaster, LA1 4YW, UK}

\begin{abstract}
McRunjob is a powerful grid workflow manager used to manage the
generation of large numbers of production processing jobs in High Energy
Physics. In use at both the DZero and CMS experiments, McRunjob has
been used to manage large Monte Carlo production processing since 1999 and
is being extended to uses in regular production processing for analysis and
reconstruction. Described at CHEP 2001, McRunjob converts core metadata
into jobs submittable in a variety of environments. The powerful core
metadata description language includes methods for converting the metadata
into persistent forms, job descriptions, multi-step workflows, and data
provenance information. The language features allow for structure in the
metadata by including full expressions, namespaces, functional dependencies,
site specific parameters in a grid environment, and ontological definitions. It
also has simple control structures for parallelization of large jobs. McRunjob
features a modular design which allows for easy expansion to new job
description languages or new application level tasks.  
\end{abstract}

\maketitle

\thispagestyle{fancy}


\section{Introduction}

  McRunjob (Monte Carlo Run Job) 
was first created in the context of the DZero Experiment at 
Fermilab during the 1999 DZero Monte Carlo Challenge.  At the time, 
there was no easy generic way to organize large batches of Monte Carlo
jobs, each possibly involving multiple processing steps. McRunjob was
originally designed so as to be generic enough so that the addition of 
new production processing executables would not pose a significant 
integration problem into the existing framework and so that different
executables could be linked together in possibly complex tree-like workflows
in which each node represents a processing step.  The main focus of 
McRunjob provides a metadata based abstraction of
each job step and to provide tools that allow for specification of the 
metadata, functional dependencies of the metadata among distinct steps, 
delegation of methods to build and or run jobs, 
and linkages to external frameworks, databases, or servers.  While McRunjob
has been used continuously at DZero since then, it has only been in use at 
CMS since the end of 2002 for regular production operations.  

  Typically, McRunjob operates during the job building stage to turn 
structured metadata into jobs.  It does this by establishing interfaces to 
do the following: 
\begin{itemize}
\item Define and access a unit of schema called a Configurator
\item Register functions to the schema to perform job building, or 
\item Optional delegation of job building responsibilities to other Configurators
\item Support User driven framework operation
\item Support linkages to external databases, catalogs, or resource brokers.
\item Register parsers to the schema to allow for customized access to the 
Configurator interface as text macros
\item Specify dependencies among the metadata elements
\item Support rudimentary ontologies through specification of synonyms and 
versioning
\item Support inter-Configurator communication and User Interface through
a Configurator container object known as the Linker.
\end{itemize}

\section{Architecture of McRunjob}

McRunjob is implemented in Python and consists of three major components: 
\begin{itemize}
\item \emph{The Configurator}
Configurators are essentially packages of metadata that describe 
applications.  Configurators can be defined to describe application input, 
environment, and output. However, since the Configurators are completely
generic, they can also describe batch queues, grid execution environments, 
information from a database, local computing site information, etc. 
Taken together, the Configurators describe workflow and provenance of
data.
\item \emph{The Script Generator}
The Script Generator is a specialization of a Configurator that also 
implements the ScriptGen interface.  The ScriptGen interface makes it 
possible for Configurators to delegate specific job generating tasks to 
a single common ScriptGen object.  This helps keep job generation consistent
in an environment where there may be different schemes for creating or
handling jobs.
\item \emph{The Linker}
The Linker is a container for Configurators.  It also acts as a 
communication bus for Configurators, a driver for the job building 
framework, and a user interface to the Linker and Configurator APIs. 
\end{itemize} 

\begin{figure*}[t]
\centering
\includegraphics{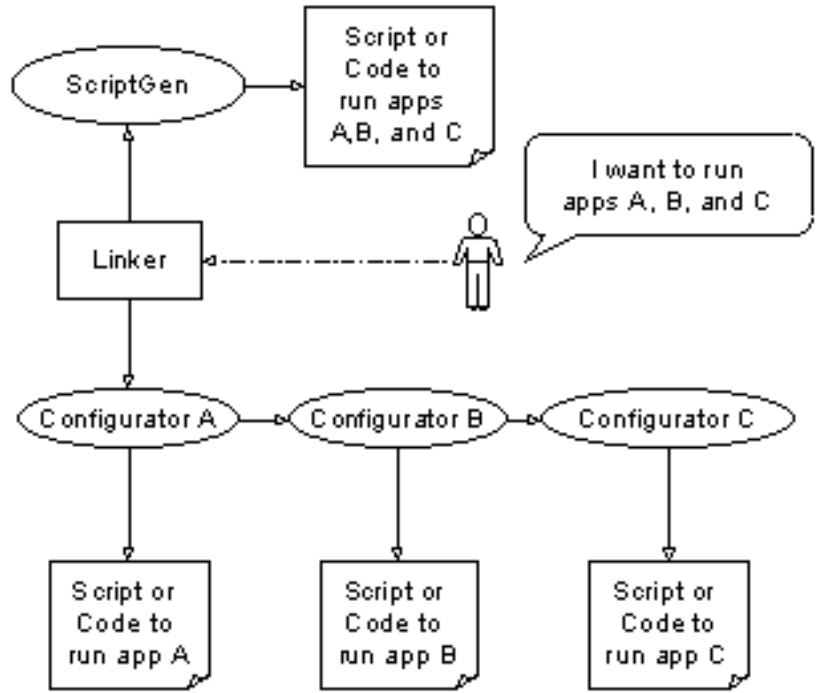}
\caption{ A simple McRunjob scenario.  The User or Production coordinator needs
to run three applications A, B, and C.  The user communicates with the Linker
to attach the appropriate configurators, set their metadata values, and 
run the Linker framework to cause the Configurators to produce jobs.} \label{fg:arch}
\end{figure*}

Figure \ref{fg:arch} shows the simplest McRunjob scenario.  A User or Production 
Coordinator needs to run three applications: let's call them A, B, and C.  Let's say
further that the output of A is the input of B and that the output of B is the
input to C.  The user will communicate to the Linker directives to instantiate 
pre-defined Configurators corresponding to A, B, and C.\footnote{This is the most common
case: that the Configurators corresponding to production applications 
are written beforehand by experts.  However, directives
also exist for the creation of Configurators and specification of schema "on the fly."}  
Usually such job building directives are kept in an McRunjob macro script, the syntax of which 
is described below.
The user issues a set of configuration macro commands which are routed to 
the relevant Configurators.  These configuration commands may include specification
of values for the schema, specification of inter-Configurator dependencies, and the 
specification of functional dependencies among schema elements in different Configurators.
Since each Configurator is required to have a unique description within the Linker space, 
so the Configurators themselves function much like namespaces.  An example of a simple 
functional dependency is B:InputFile = A:OutputFile.\footnote{Such obvious I/O dependencies 
have a special place in many job handling systems, but McRunjob treats all possible 
metadata dependencies on an equal footing.}  The ``MakeJob'' and ``MakeScript'' directives,  
examples of framework calls, are issues.  These particular framework call cause the Configurators 
to generate shell scripts to handle their respective applications in serial order.  
The scripts are then collected by the Linker and a composite shell script that represents the 
entire workflow is produced.  This procedure can be reset and re-run as many times as 
desired to kick out as many jobs as desired.  The procedure is also generic in that different
targets than shell scripts (eg- directed graphs) can be selected by including different
ScriptGen modules.

\begin{figure*}[t]
\centering
\includegraphics[width=150mm]{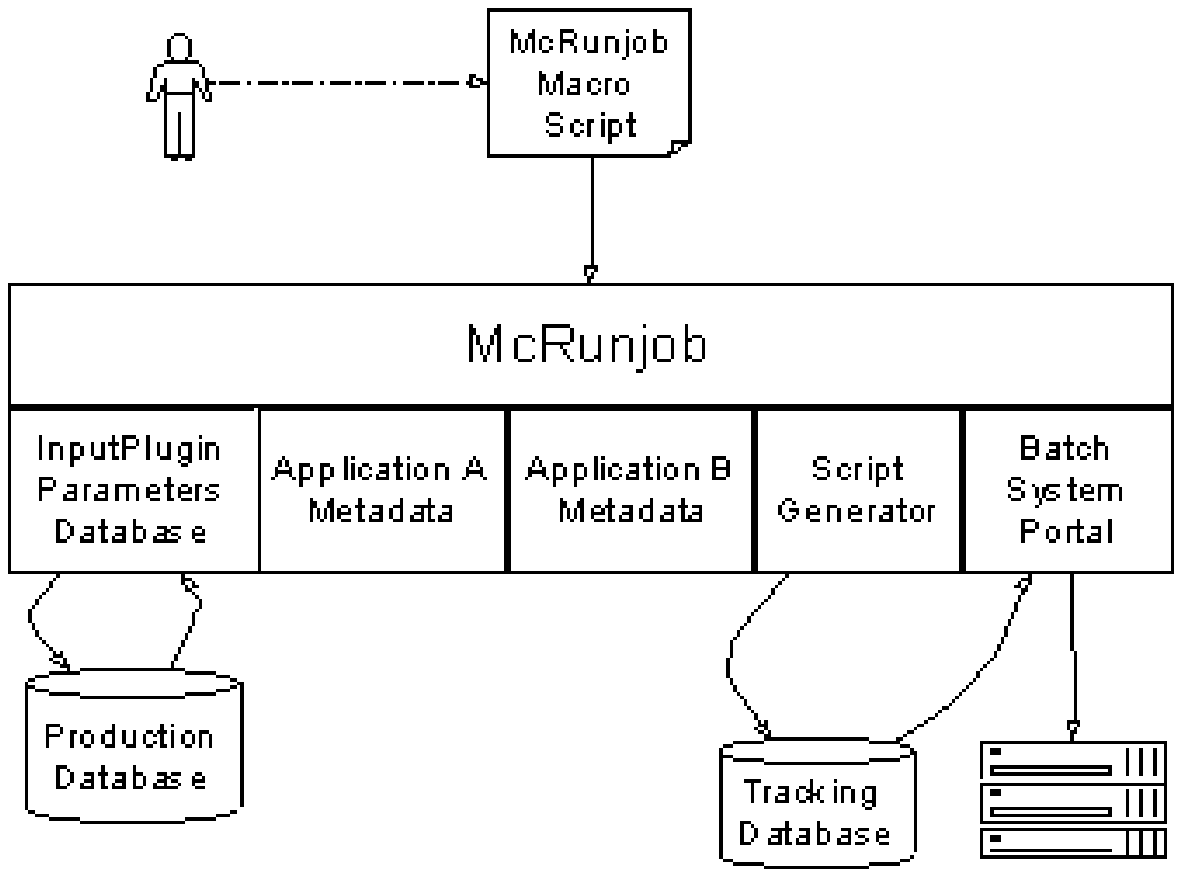}
\caption{ At the direction of the user through a macro script file, one
class of Configurators ("InputPlugins") have backends that communicate 
to external databases, planners, or servers.  More conventional Configurators just hold on the 
application metadata. ScriptGenerators collect results from previous Configurators and
produce composite workflows as described above.  Finally, a Batch Portal Configurator may 
take the produced composite script object and submit it to a batch queue.} \label{fg:outside}
\end{figure*}

In addition to modeling the application space, the Configurators also provide a useful 
abstraction through which to 
exchange information with other external sources such as databases, batch queues, etc. 
Figure \ref{fg:outside} shows a generalized picture of how Configurators may do this.  Typically, the 
user writes a script of McRunjob macro commands which are interpreted by the Linker 
framework (shown in light blue.)  The Linker takes these commands and distributes them to 
the Configurators attached below.  The Configurator layer exposes to the Linker sets of 
metadata key-value pairs, but with additional customizable backends.  For example, one
class of Configurators ("InputPlugins") have backends that communicate 
to external databases, planners, or servers.  More conventional Configurators just hold on the 
application metadata. ScriptGenerators collect results from previous Configurators and
produce composite workflows (as described above).  Finally, a Batch Portal Configurator may 
take the produced composite script object and submit it to a batch queue.

In the DZero context, Monte Carlo production is coordinated with the SAM database at 
FNAL.  Two of the common applications in the workflow are PYTHIA Generation and D0gstar 
(GEANT Simulation.) The ScriptGenerator targets executable scripts for the DZero executable
script environment\footnote{There is only one ScriptGen at DZero so usually no distinction is made.}, 
and one possible execution environment is the SAM/JIM grid service. 
In upcoming DZero production on the grid, there may no jobs; rather the focus is on 
automatic production of McRunjob macros which replace scripts and are executable by remote 
Linkers.  Also, there is work being done to leverage existing 
McRunjob tools to do monitoring on the DZero farms.

Some typical dependency relationships among configurators include modeling of the 
sequence in which applications have to run on a set of events in order to reach a given 
data product or modeling of parameter flows in environments where several databases or
configuration files may be consulted in the process of job creation.  One feature
of the McRunjob framework in CMS that is disabled in the DZero framework is the requirement that 
such dependency relationships be clearly defined before inter-Configurator parameter lookup can
take place.  This discipline is useful, however, in an environment where a clear provenance 
of the produced data is not already established by central means.  At DZero, this is largely 
handled by the SAM database.

Three final points can be made.  The first is that although McRunjob was conceived in a
Monte Carlo production environment, it is perfectly and immediately well suited to any
problem involving complex workflow specification and job templating in a production 
processing environment.  The second is that while McRunjob was designed to describe
production workflows in the Monte Carlo setting (ie- applications and files) there is
no reason that it cannot be extended into more fine grained settings to describe Analysis
Object Data (AOD) and their relationships and provenances.  Finally, McRunjob typically
operates after metadata is specified and before jobs are actually submitted; McRunjob 
could conceivably be extended into runtime to bring parameter lookup services into
runtime.  

\subsection{The Configurator}

The Configurator API provides methods for automating many of the procedures 
inherent in specifying workflow for Monte Carlo Production or Analysis.
The Configurator is essentially a value added metadata container. It comprises
a special TriggerDictionary class used to hold the metadata key/value pairs and
the methods provided to manipulate the metadata in a production processing 
environment.

\begin{itemize}
\item The TriggerDictionary allows the user to provide an implementation
for the internal dictionary.  The implementation must use the regular
Python UserDict interface.
\item The TriggerDictionary makes calls to user supplied functions on reads
or writes to the internal dictionary implementation. 
\end{itemize}

The TriggerDictionary triggering mechanism is used to implement several 
Configurator functionalities, such as parameter lookup or construction.  It 
is also used to implement parameter monitor and watch functions for debugging 
purposes.  The internal implementation object is swappable, enabling GUI
linkage on demand.  There are four kinds of triggers: (1) Global Read: 
Functions that are called when any element is read.  (2) Global Write: 
Functions that are called when any element is written to.  (3) Indexed Read: 
Functions that are called only when a specific element is read. 
(4) Indexed Write: Functions that are called only when a specific element is 
written to.  Functions that handle any of the triggers must be registered 
to the TriggerDictionary object as described below, and must accept a Python list 
as argument. In all cases, the first element in the list is always a back 
reference to the TriggerDictionary object, and the second is always the 
key that was called. The remaining elements, is present, are defined at 
registration time. NOTE: Trigger handlers registered to TriggerDictionary, 
if they are going to alter dictionary state, must always interact with the 
TriggerDictionary using the UntriggeredRead and UntriggeredWrite methods; 
otherwise an infinite loop could occur.  

The feature that the TriggerDictionary can accept any conformant implementation 
of its internal dictionary structure implies that structures can be built for this
purpose that have external linkage to graphics or GUI packages.  Furthermore, these
can be ``HotSwapped'' so that graphics packages or debugging mechanisms can be inserted 
into running McRunjob programs.

 Configurators are themselves described by metadata. 
This metadata is used internally by McRunjob to resolve dependencies, keep 
track of schema versions, resolve entries in synonym tables, and distinguish
Configurators within the memory space of a Linker. Configurators can function
within the Linker as namespaces; the ConfiguratorDescription objects allow
the namespaces to be referenced.

The ConfiguratorDescriptions are generally used internally for two things: to 
implement inter-Configurator dependencies and to aid in parameter lookup.  
In the first capacity,
a Configurator can declare dependencies on other Configurators.  This 
can occur \emph{statically} when a developer is modeling underlying 
relationships among applications or \emph{dynamically} when a user is
modeling relationships among servers, planners or databases.  When adding
a Configurator to the Linker or when altering the dependencies of 
Configurators already in the Linker, the dependencies are checked and 
an Exception is thrown if not satisfied.  The mechanism is that the 
dependencies of the new or changed Configurator are matched 
against the list of existing Configurators in the Linker
If there is not a match, then an exception is
thrown.  NOTE: This behavior is disabled in DZero. 

The Configurators support a parameter lookup service based on namespaces within the 
Linker and on declared dependencies.  Since the ConfiguratorDescriptions of Configurators
must be unique within the Linker, they define a partition (namespaces) on the parameters.
Thus a parameter in the Linker 
is defined by a complete specification of the ConfiguratorDescription
and the parameter name.  From a Configurator point of view, a parameter in Configurator B
is only visible if there exists a declared dependency on Configurator B.  This last 
behavior is also disabled in DZero.  

Configurators contain synonym tables.  These are lookup tables that 
translate local metadata key names into different key names in other Configurator
types.  The behavior of a workflow can therefore change depending upon what 
synonyms are loaded at any given time.  The synonyms tables can be loaded for 
different environments or changing versions, thus providing for a rudimentary 
ontology.

Finally, Configurators can have explicit metadata translation or construction rules
attached directly to each metadata element.  These are available to the developer, but not yet
available in the macro script language.

Examples of Configurators include those that have connectivity with external 
databases (ie- RefDB in CMS through SQL queries or SAM in DZero through system
commands,) those which model applications steps (ie- Monte Carlo generation, 
detector simulation, digitization,) those which submit jobs to specified batch
portals (ie- LSF or PBS batch systems, Condor, DAGMan/Condor-G.) 

\subsection{The Script Generator}

One of the problems encountered in practice using the above model of 
Configurators generating custom bits of code which are then collected by 
the Linker for submission to an execution manager is that there is no
organization in place to help guarantee that all of the independently generated
bits of code will be compatible.  For example, they may be targeted for an 
environment in which the code bits cooperate at 
runtime through non-McRunjob interfaces.  The only way to 
organize this is at the level of the Configurator itself; so the number
of modules potentially needing modification in case of a change to the
runtime environment is as large as the number of Configurators. 

ScriptGen is a special interface implemented by some Configurators that enable 
Configurators to delegate specific calls to a single Configurator.  In the case
of delegation, the ScriptGen must declare ConfiguratorDescriptions and method 
calls which it can handle.  The Configurator must specify which method calls it 
will delegate and the description of the ScriptGen module to which it is delegating.
With this functionality, a new way to organize code is available: all of the 
script generating code targeting specific runtime environments can be collected in
a single ScriptGen module.  The ScriptGen module is also usually the agent which
the Linker uses to collect code bits targeted for a specific environment in order
to create a composite job or DAG.

  Examples of different ScriptGen modules in CMS are the default ImpalaScriptGen module,
which generates executable scripts compatible with the legacy Impala production 
environment, the ImpalaLiteScriptGen module, the CMSProdScriptGen module, the 
VDLScriptGen module for generating specifications written in the Chimera Virtual Data
Language, and the MOPDagGen module for taking the output of other specified 
ScriptGen modules and producing a Directed Acyclic Graph (DAG) for use by the 
Condor DAGMan tool.  

\subsection{The Linker}

The Linker is a Container class for Configurators.  It handles all communication between the 
User and the Configurators and between any two Configurators.  It also contains a repository
for ``script Objects''.  Configurators that need to generate code bits to implement a 
given workflow or job can store these bits in the Linker as script Objects.  As described
above, a ScriptGen module may later collect script Objects targeted for a specific environment 
and create a composite script Object.  It may also, as in the case of MOPDagGen, wrap 
existing scriptObjects or composites into a DAG.  

The Linker also supports some simple looping structures within the McRunjob macro scripts, 
and also drives the framework, described in the next section.  

\section{The Framework}

The Configurators build jobs together by contributing their 
specialized knowledge of application steps or external resources 
to the overall whole in structured ways.  One part of this 
structure is the Configurator dependencies\footnote{Or, when not enabled, just the order 
in which Configurators are added.}.  Another structure which 
organizes the order in which tasks are completed is the Framework.
The framework is basically a sequence of strings used as messages sent to 
framework handlers in the Configurators. The messages can include things 
like Reset, MakeJob or MakeScript for shell script building, 
listing of derivations and transformations in CHIMERA, etc.  

Traditionally in McRunjob, framework calls are handled directly by the 
Configurators themselves through subclassing the Framework handling methods.  
However, to better support flexibility without using inheritance, the Configurator
base class also provides methods for registering functions (possibly user supplied 
in certain simple cases) to handle specific framework messages. As described above, 
as a double indirection supporting code maintenance tasks, these functions can  
also be registered to a special Configurator that inherits the ScriptGen interface
and then delegated.

The Linker thus provides the drumbeat according to which the Configurators march: it 
provides a context within which to order the Configurators by their dependencies and
a framework within which to sequence method invocations.

\section{The Macro Language}

The McRunjob macros are intended to provide a user interface to the Configurator and Linker APIs.  
It is possible to construct the macros as a complete declarative specification of the 
workflow, but even in a procedural environment where parameters are being ``constructed'' or
``discovered'' in external databases the resulting state of the McRunjob program can at any 
time be dumped in declarative format.  Thus  
macros can also serve as a rudimentary "provenance" for the described or constructed 
workflow.  

  The Linker macros comprise commands that attach Configurators, route macro commands
to specified Configurators, and simple looping and conditional constructs. In the Configurator, 
the handling of macros is done in a ``class distributed'' fashion. Configurator classes can have 
macro handlers registered to them so that it is very easy to extend their macro interfaces.
A particular Configurator object passes a particular macro to each of the registered
macro handlers until it finds one that can handle the particular macro.  The Configurator base 
class registers a base parser which is called last, and Configurator subclasses extend this.
Following is a list of simple Linker directives: 
\begin{itemize}
\item \textbf{attach cfgIdentifier} attaches a configurator of the given type.
\item \textbf{cfg cfgIdentifier cmd} issues the macro ``cmd'' to the specified Configurator.
\item \textbf{framework run cmd} issues the framework message ``cmd'' to all Configurators in 
sequence.  Framework commands can be grouped together and run in groups as well.
\end{itemize}
Following is a list of simple Configurator macros:
\begin{itemize}
\item \textbf{additem keyname}  Adds a metadata element named ``keyname''
\item \textbf{define keyname expression}  Sets the value of ``keyname'' to ``expression'' where
expression can be a literal or a reference to the value of a key in another configurator or a
reference into the internal Configurator synonym table\footnote{`Real expressions like ``a+b/c'' are 
not yet supported. } or a directive to construct the value by registered function.
\item \textbf{addreq cfgIdentifier}  Adds cfgIdentifier as a dynamic 
dependency for this configurator.
\item \textbf{synonym key ::cfgIdentifier:newkey}  Defines a possible synonym  for ``key'' to 
target ``newkey'' in another Configurator.
\item \textbf{oncall fmk do cmd} Store command ``cmd'' and execute it on receipt of framework 
call ``fmk''.
\end{itemize}

\begin{table*}[t]
\begin{center}
\begin{tabular}{|c|c|c|c|c|c|}
\hline FrameworkCall & \textbf{HelloWorld} & \textbf{HelloWorld} & \textbf{HelloWorld} 
& \textbf{ScriptGen} & \textbf{Fork} 
\\
\hline 
  Reset & Handled & Handled & Handled & Handled & Handled \\
\hline
  MakeJob & Delegated & Delegated & Delegated & Skipped & Skipped \\
          & to ScriptGen & to ScriptGen & to ScriptGen &         &         \\
\hline
  MakeScript & Skipped & Skipped & Skipped & Handled & Skipped \\
\hline
  RunJob & Skipped & Skipped & Skipped & Skipped & Handled \\
\hline
\end{tabular}
\label{tb:framework}
\caption{Framework Operation in the Hello World example.  The sequence goes from 
left to right and then up to down, like reading a book (in English.)}
\end{center}
\end{table*}

  Macros can source other macros.  In this way, McRunjob macro commands can be separated into
synonym definitions and stored commands on one hand and pure workflow descriptions on the other hand.
The former are seen as part of the environment and are in some sense independent of the pure 
workflow descriptions.  The management of these environments leads to a rudimentary ontological 
management system.

\subsection{The ``Hello World'' Example}

In the CMS implementation of McRunjob, a HelloWorld example is provided which consists of 
a HelloWorld Configurator with metadata element HelloMessage and a HelloWorldScriptGen that 
also serves as a metadata server.  Each HelloWorld configurator is equipped to produce a 
short script which echos its HelloMessage to the screen.  The HelloWorldScriptGen collects
these scripts into a composite.  The following is a simple example macro script fragment
that would print out a HelloWorld message in English, French, and German\footnote{This uses new 
syntax instituted as of May 2003.}.

\begin{verbatim}
# Attach the ScriptGen which will in this 
# case also serve metadata values to the 
# HelloWorld configurators
attach HelloWorldScriptGen
cfg HelloWorldScriptGen additem English
cfg HelloWorldScriptGen define English \
                               Hello World
cfg HelloWorldScriptGen additem French
cfg HelloWorldScriptGen define French \
                            Salut le Monde
cfg HelloWorldScriptGen additem German
cfg HelloWorldScriptGen define German \
                            Hallo Welt

# Attach the HelloWorld Configurators 
# themselves
attach HelloWorld named English
attach HelloWorld named French
attach HelloWorld named German

# Enable HelloWorld to delegate script 
# generation to ScriptGen. (This also 
# sets correct dependencies.)
cfg HelloWorldScriptGen register HelloWorld 
      
# Route the metadata to correct 
# configurators
cfg HelloWorld named English define \
 HelloMessage ::HelloWorldScriptGen:English
cfg HelloWorld named French define \
 HelloMessage ::HelloWorldScriptGen:French
cfg HelloWorld named German define \
 HelloMessage ::HelloWorldScriptGen:German

# Fork the resulting jobs in background
# Set it to get executables list every time
# ``RunJob'' is executed.
attach Fork
cfg Fork define ScriptGenName \
              HelloWorldScriptGen
cfg Fork oncall RunJob do \
      define ExecutableList ::construct

\end{verbatim}

Upon invocation of the framework, this will result in the sequence of
framework calls shows in table \ref{tb:framework} and will result in the
output 

\begin{verbatim}
Hello World
Salut le Monde
Hallo Welt
\end{verbatim}

\section{Conclusions and Future Plans} 

McRunjob has been successfully used in both the DZero and CMS experiments to 
model HEP workflows for Monte Carlo productions both on local controlled farms 
resources and in Grid environments.  In both experiments, there is a desire to 
see how far we can extend McRunjob into the realm of interactive analysis;
The extension to batch analysis should be straightforward.  More immediately, 
full expression support will be added to the macro language.  A common project
at Fermilab between USCMS and DZero is also being started to  address 
common goals and support issues.

There are many exciting directions being explored.  In the context of DZero, 
runtime McRunjob is being explored as an answer to the need for monitoring 
jobs on the farms.  The declarative specifications of jobs are converted to 
XML and stored in a local XML database, and the McRunjob created job is instrumented
to update this database.  Furthermore, the extension of the rudimentary ontologies as
described above presents an interesting research problem as the environments (as defined
above) become large.  Also, how the workflow description plus and annotations from the 
environment informs the provenance of a particular derived data product is an open
question.  Finally, as the Grid itself adopts a more Web Services oriented model of
operation, it may become important to include extensions to proposed standards 
such as Web Services Flow Language (WSFL). 

\begin{acknowledgments}
The authors wish to thank the members of the DZero and CMS experiments who
have provided many insights (and bug reports) over the years; especially 
Boaz Klima, Kors Bos, Willem van Leeuwen, Lee Lueking, and the SAM team at DZero; 
and Tony Wildish, Veronique Lefebure, and Julia Andreeva at CMS; and Jaideep 
Srivastava and Praveen Venkata of the University of Minnesota, Peter Couvares, 
Alan DeSmet, and Miron Livny of the University of Wisconsin, and Richard Cavanaugh
and Adam Arbree of the University of Florida, and Muhammad Anzar Afaq of Fermilab
for helpful discussions.
\end{acknowledgments}


\begin{thebibliography}{9}   


\bibitem{bb:gg00}
G.E. Graham, ``DZero Monte Carlo'' ACAT'00, Fermilab, Batavia, IL, USA 2000

\bibitem{bb:gg01a}
D. Evans and G.E. Graham, ``DZero Monte Carlo Production Tools (8-0127)'' CHEP'01, Beijing, China 2001

\bibitem{bb:gg01b}
G.E. Graham, T. Wildish, et al. ``Tools and Infrastructure for CMS Distributed Production (4-033)'' CHEP'01, Beijing, China 2001

\bibitem{bb:dz}
DZero McRunjob page \textbf{http://www-clued0.fnal.gov/runjob}

\bibitem{bb:cms}
USCMS McRunjob page \textbf{http://www.uscms.org/scpages/subsystems /DPE/Projects/MCRunjob}

\end{thebibliography}
\end{document}